# Are Crab Nanoshots Schwinger Sparks?

**Albert Stebbins**
Fermilab

**Hojin Yoo**
University of Wisconsin, Madison
Fermilab
Theory Group, Lawrence Berkeley Laboratory
Berkeley Center for Theoretical Physics, University of California, Berkeley

## Abstract

The highest brightness temperature ever observed are from "nanoshots" from the Crab pulsar which we argue could be the signature of bursts of vacuum $e^\pm$ pair production. If so this would be the first time the astronomical Schwinger effect has been observed. These "Schwinger sparks" would be an intermittent but extremely powerful, $\sim 10^3\, L_\odot$, 10 PeV $e^\pm$ accelerator in the heart of the Crab. These nanosecond duration sparks are generated in a volume less than $1\,\mathrm{m}^3$ and the existence of such sparks has implications for the small scale structure of the magnetic field of young pulsars such as the Crab. This mechanism may also play a role in producing other enigmatic bright short radio transients such as fast radio bursts.

May 21, 2015   submitted to *The Physical Review Letters*

## Introduction

Giant pulses (GPs) has been one of the most interesting phenomena from pulsars since their first observation from the Crab pulsar in 1968 [1]. The properties of GPs have been explored for many years, and may provide important clues to pulsar emission mechanism. In particular, Hankins *et al.* [2] and more recently Jessner *et al.* [3] reported nanosecond-long sub-pulses within GPs from the Crab, so called "nanoshots". Discounting the possibility of strong relativistic beaming toward us we can infer a brightness temperature at emission as high as $10^{38}\,\mathrm{K}$ ($\sim 10^6\,T_\mathrm{Planck}$) and the size of the emission region at 30cm or smaller. By these measures the source of these nanoshots are the brightest objects ever observed in the universe as well as the smallest objects ever detected outside our solar system. This extreme pheonoma may play an important role in the overall energetics and evolution of young pulsars such as the Crab.

In this Letter, we propose a novel theoretical model for nanoshots from the Schwinger effect. We will first go through the numerology of what would be required to produce nanoshots. Then we show this to be consistent with electomagnetic pulse (EMP) from bursts of vacuum pair production (Schwinger effect) giving a simple formalism for the electromagnetic pulse of ultra-relativistic Schwinger sparks in a strong magnetic field. Various physical quantities are derived from the formalism, and we discuss the properties of the EMP from Schwinger sparks as a model for nanoshots. The pairs produced will be accelerated to very high-energies due to the strong electric field ( > 10 PeV), which potentially may be of interest for BSM and neutrino physics.

## Nanoshots

Jessner *et al.* [3] have detected unresolved individual electromagnetic pulses with a peak flux up to $S_\nu = 150\,\mathrm{kJy}$ in a band $\nu = 8.5 \pm 0.2\,\mathrm{GHz}$, although typical fluxes presented in that paper are $S_\nu \sim 10\,\mathrm{kJy}$. Whatever mechanism produces these pulses must be able to produce the largest pulse which is what we consider in this section. The Crab pulsar is known to be at a distance $D_\mathrm{crab} = 2.2\,\mathrm{kpc}$ from Earth [4] so this corresponds to a radio bolometric luminosity of $\dot{\mathcal{E}} \approx 4\pi D_\mathrm{crab}^2\, S_\nu\, \delta\nu \gtrsim 10^2\, L_\odot$. The radio pulses have duration $\delta t \lesssim 1\,\mathrm{ns}$ suggesting the diameter of the emission volume to be $L \lesssim \delta t/c \approx 30\,\mathrm{cm}$ and indicating a peak brightness temperature

$$kT_\nu^{\text{peak}} \approx \frac{1}{2}\left(\frac{c}{\nu}\right)^2 S_\nu \left(\frac{D_{\text{crab}}}{2L}\right)^2 \gtrsim 2\times 10^{24} \text{ erg}. \tag{1}$$

This number, millions of times the Planck energy, requires macroscopic coherent emission by large number of charged particles moving in synchrony rather than a sum of microscopic single particle emission. At emission the propagating electric field is $\delta\text{E} \approx \sqrt{\frac{8\pi \dot{\mathcal{E}} \delta t}{L^3}} \gtrsim 6\times 10^{11}$ Gauss, which in magnitude is similar to the "typical" magnetic field expected at the surface of neutron stars. Neutron star magnetic fields are known to vary by many orders of magnitudes but an often quoted estimate for the peak magnetic field of the surface of the Crab pulsar is $B_{\text{crab}} \approx 3\times 10^{13}$ Gauss. The electric current needed to produce this field corresponds to moving a charge $Q \approx \frac{1}{4\pi} L^2 \delta\text{E} \approx 10^4$ Coulomb (0.15 mole of $e^\pm$) across 30 cm at the speed of light, or if smaller velocities are involved then even greater charge is required. What is surprising is the large amount of charge in such a small volume. Typical pulsar magnetosphere models assume a Goldreich-Julian density [5], which for the Crab is $n = B_{\text{crab}}/(c\, e\, P_{\text{crab}}) \lesssim 10^{-3}$ mole$/$m$^3$ ($P_{\text{crab}} = 33$ msec is the Crab rotational period). With this starting density one must produce $\sim 1$ mole$/$nsec$/$m$^3$ *in situ* which would be difficult with a normal pair cascade.

The similarity in magnitude of $\delta\text{E}$ and $B_{\text{crab}}$ itself is suggestive that nanoshots are generated near the neutron star surface and related to transients in the electromagnetic (EM) field of the neutron star. While short transients in the EM field are not part of normal pulsar modelling it would not be surprising to find this phenomena in young pulsars such as the Crab. Highly magnetized and rapidly rotating neutron stars are hugely out of thermodynamic equilibrium with their environment and will utilize all available channels to equilibrate; shedding energy, angular momentum and magnetic field. The initial magnetic field frozen into the neutron star material could be quite convoluted with large field variations on fairly small scales since high multipole fields do not contribute significantly to rotational energy loss and are unconstrained by observations. Small $B$-field loops may be shaken or if loosely tethered even become detached. Similar surface phenomena power stellar flares and soft gamma repeaters which occur in magnetars (neutron stars with $B > E_S$).

### QED and the Schwinger Limiting Field

The Crab pulsar's magnetic field is large enough to put it close to if not in the magnetar classification ($B \gtrsim E_S = \frac{m_e^2 c^3}{e\hbar} = 4.41\times 10^{13}$ Gauss) even if only in localized regions ("hot spots") on its surface. It is well known that the large $B$ near the surface of magnetars lead to various "exotic" QED phenomena such as photon splitting [6]. The fact that the inferred electric field, $\delta\text{E}$, of the nanoshots at their source are also close to $E_S$ suggests that the nanoshots themselves could be a QED phenomenon. Schwinger [7] computed in QED the rate of production of $e^\pm$ pairs in a strong electric field. This is a threshold phenomena which turns on rapidly as the electric field, $E$, approaches the Schwinger critical field $E_S$ defined above. $E_S$ is often called the "limiting field" but as we shall see the actual limiting field is significantly smaller and closer to the nanoshot $\delta\text{E}$. The relevant electric and magnetic field strength, $E$ and $B$, is that measured in a colinear frame where $\boldsymbol{E} \parallel \boldsymbol{B}$. Near the surface of a neutron star the EM field we expect $B \gg E$ and if $E \ll E_S$ (it only has to be down by a factor of a few) then the pair production rate (pairs per unit volume per unit time) is given by [8] $\Gamma[E] \cong \frac{e^2}{\pi \hbar c} B E\, e^{-\pi \frac{E_S}{E}}$. In this case nearly all pairs are created in the lowest Landau level with their spins aligned to minimize the energy. The pairs separate and move in opposite directions so as to short out the $E$ field. If the electric field is coherent over a macroscopic volume then many pairs can be created and it will take a light crossing time for them to separate.

In the MHD approximation, used to model long timescale phenomena in a pulsar magnetosphere, $\boldsymbol{E}\cdot\boldsymbol{B} = 0$ so $E = 0$ and hence $\Gamma = 0$. However for short transient phenomena such as magnetic field reconnection one can expect the MHD approximation to be violated. The Schwinger mechanism is one way of limiting or shorting the fields so as to bring it back to a MHD state. Note that for $E < E_S$ that the pair production rate is an extremely rapidly increasing function of $E$. The way the Schwinger mechanism shorts out an increasing $E$ field is that initially nothing much happens as $\Gamma$ is too small to effect $E$. Eventually the field reaches a limiting value, $E_{\text{lim}}$, at which point the pair current suddenly overwhelms the increasing $E$ and the field never increases much beyond $E_{\text{lim}}$. It is assumed the transient is such that the field would eventually relax to zero even without shorting, so the Schwinger effect merely limits the magnitude of the electric field excursion. Any episode of pair creation, which we call a *Schwinger spark*. Eventually the pairs are expelled from the region in oppositely directed outgoing bunches of $e^+$'s and $e^-$'s. As the charges separate they will produce an outgoing EMP which we propose is the observed nanoshot. If the field excursion is over a large enough volume (not the case considered here) then the pair current can actually reverse the field creating a persistent pair plasma and $E$ will oscillate with a decreasing amplitude about zero [8].

To model a Schwinger spark we assume a uniform constant $\boldsymbol{B}$ with $E \ll B$ in the colinear frame with direction $\boldsymbol{B} \parallel \boldsymbol{E} = E\hat{z}$. In cases of interest $eEL \gg m_e c^2$ so that any pairs produced are rapidly accelerated to the speed of light. We suppose the field dynamics are completely electromagnetic so the fields evolve on a light-crossing time. For ultra-relativistic pairs produced via the Schwinger effect Maxwell tells us $\ddot{E} = -4\pi \dot{J} = -8\pi e\, \Gamma[E]$ (see [8], this does not include the negligible polarization current, corresponding to the initial microscopic separation of the pairs). Pair production will limit the electric field $E$ to a limiting value, $E_{\text{lim}}$, which we assume occurs at $t = t_*$. Near this peak

$$E[t] = E_{\text{lim}}\left(1 - \tfrac{1}{2}\left(\tfrac{t-t_*}{\delta t_E}\right)^2 + \ldots\right) \qquad \delta t_E \equiv \sqrt{\frac{E_{\text{lim}}}{8\pi c\, e\, \Gamma[E_{\text{lim}}]}}$$

$$\ln[\Gamma] = \ln[\Gamma[E_{\text{lim}}]] - \tfrac{1}{2}\left(\tfrac{t-t_*}{\delta t_\Gamma}\right)^2 + \ldots \qquad \delta t_\Gamma \equiv \frac{\delta t_E}{\sqrt{\pi \frac{E_S}{E_{\text{lim}}} + 1}} \tag{2}$$

Solving for $E_{\text{lim}}$ in terms of $\delta t_E$ we find

$$E_{\text{lim}} = \frac{\pi E_S}{\ln\left[\frac{8 e^3\, \delta t_E^2\, B}{\hbar^2}\right]} = \frac{2.75 \times 10^{12}\ \text{Gauss}}{1 + 0.02\, \ln\left[\left(\frac{\delta t_E}{\text{nsec}}\right)^2 \frac{B}{10^{13}\ \text{Gauss}}\right]}, \tag{3}$$

which is over an order of magnitude below the Schwinger critical field, $E_S$. Note the extremely weak dependence on the parameters $\delta t_E$ and $B$. A similar formula with nearly the same logarithm can be derived even in absence of a magnetic field. This robust prediction for a limiting field of $E_{\text{lim}} \approx 2.7 \times 10^{12}$ Gauss $\cong 7.5 \times 10^{16}\ V/\text{m}$ in almost any macroscopic context is our most far-reaching result, and corrects downward the naive estimate of $E_S$ for the limiting field. The closeness of $E_{\text{lim}}$ to the nanoshot source electric field, $\delta E$, computed for the brightest nanoshot is a clue that the nanoshots may be related to the Schwinger mechanism. We find that, apart from the very weak parameter dependence, $\delta t_\Gamma \cong 0.14\, \delta t_E$. The pair production rate is well approximated by a Gaussian, $\Gamma \approx \Gamma[E_{\text{lim}}] \operatorname{Exp}\!\left[-\frac{(t-t_*)^2}{2\, \delta t_\Gamma^2}\right]$, which has the rapid turn-on/off behaviour. The density of pairs produced in the spark is

$$n_{\text{pair}} = \int \Gamma[E(t)]\, dt \approx \frac{E_{\text{lim}}}{4\sqrt{2\pi}\left(\pi \frac{E_S}{E_{\text{lim}}} + 1\right) c\, e\, \delta t_\Gamma} \cong 0.61\, \frac{\text{nsec}}{\delta t_\Gamma}\, \text{mole}/\text{m}^3 \tag{4}$$

which provides the charge required to produce the EMP corresponding to a nanoshot. These results are for uniform fields, but we have found similar numerology with 1D Maxwell-Vlasov simulations. Nanoshots must be generated from 3D sparks which we expect have similar numerology.

### Schwinger ElectroMagnetic Pulse

Here we give a summary of EMP emission from a Schwinger spark (see [9] for more details). A single bunch of charged particles moving at constant velocity does not emit EM radiation. A spark consists of two bunches, electrons and positrons, moving at nearly constant velocity $\cong c$, but in opposite direction. As the two bunches separate they will produce a large EMP, which we call *charge separation EMP*, which for $\nu\, \delta t \lesssim 1$ has power $\propto N_{\text{pair}}^2$. Particles in each bunch will be accelerating, getting closer to the speed of light, but the Larmor radiation from this acceleration is small compared to the pulse from charge separation although Larmor radiation will extend to larger frequency. For larger $\nu$ the EMP will be less and less coherent, until for high enough frequency the outgoing power will be incoherent, *i.e.* the sum of the power from individual particles $\propto N_{\text{pair}}$ and this is where the acceleration radiation will be most evident. We have estimated $N_{\text{pair}} \sim 10^{23}$ so incoherent emission is highly suppressed with respect to coherent emission.

The electromagnetic field from a localized charge/current distribution at the large distance $D$ can be expressed as $\boldsymbol{E} = \frac{1}{cD}\left(\hat{\boldsymbol{r}} \otimes \hat{\boldsymbol{r}} - \boldsymbol{I}\right) \cdot \dot{\boldsymbol{Q}}$ and $\boldsymbol{B} = \frac{1}{cD}\, \dot{\boldsymbol{Q}} \times \hat{\boldsymbol{r}}$ where the "current vector" is

$$\dot{Q}[\mathbf{x}, t] \equiv \frac{1}{c} \int d^3 \mathbf{x}' \, \dot{\mathbf{j}}\left[\mathbf{x}', t - \frac{|\mathbf{x} - \mathbf{x}'|}{c}\right] \,, \tag{5}$$

$\mathbf{J}$ is the charge current density, $\dot{\mathbf{J}}$ it's time derivative and $\dot{\mathbf{Q}}$ an integral over the past light cone.

In the colinear frame, charges will always move along the $B$ field so $\mathbf{J} = (J_+ + J_-)\hat{\mathbf{z}}$ where the $J_\pm$ are the $e^\pm$ currents and $\dot{J}_\pm = e\,\Gamma \mp c\,\frac{\partial}{\partial z} J_\pm$ (vacuum pair production plus ultra-relativistic motion). The key simplification is assuming that the particles always move at the speed of light, which is actually a reasonable approximation. Since ballistic motion cannot contribute to EMP we can integrate out the $\frac{\partial}{\partial z} J_\pm$ contribution to a term which depends only on $\Gamma$ obtaining

$$\dot{\mathbf{Q}} = 2\,e\,\dot{N}\,\text{Sec}[\theta]^2\,\hat{\mathbf{z}} \qquad \dot{N}[\mathbf{x}, t] = \int d^3 \mathbf{x}' \, \Gamma\!\left[E\!\left(\mathbf{x}', t - \frac{|\mathbf{x} - \mathbf{x}'|}{c}\right)\right] \tag{6}$$

where $\text{Sin}[\theta] = \hat{\mathbf{z}} \cdot \hat{\mathbf{r}}$ and $\dot{N}$ is the rate of pair production on the past light cone. $\dot{N}$ can be negative if the field reverses ($E < 0$). The outgoing EM energy flux density and spectral fluence is given by

$$\mathbf{S}[\mathbf{x}, t] \equiv \frac{c}{4\pi}(\mathbf{E}\times\mathbf{B}) = \frac{\dot{\mathbf{Q}}\cdot\dot{\mathbf{Q}} - (\hat{\mathbf{r}}\cdot\dot{\mathbf{Q}})^2}{4\pi D^2 c}\hat{\mathbf{r}} = \frac{(2\,e\,\dot{N})^2}{4\pi D^2 c}\text{Sec}[\theta]^2\,\hat{\mathbf{r}} \tag{7}$$

$$F_\nu[\mathbf{x}] = \frac{(2\,e)^2}{4\pi D^2 c} \int_{-\infty}^{\infty} dt\,\left|e^{-i\,2\pi\nu t}\,\dot{N}[\mathbf{x},t]\right|^2 \text{Sec}[\theta]^2 \tag{8}$$

Note that the $\text{Sec}[\theta]^2$ divergence at $\theta = \pm\frac{\pi}{2}$ would be regulated by relaxing the luminal velocity approximation.

At low frequencies, $F_\nu = F_0 = \frac{(2\,e\,N\,\text{Sec}[\theta])^2}{4\pi c D^2}$ where $N \equiv \int_{-\infty}^{\infty} dt \int d^3\mathbf{x}\,\Gamma[\mathbf{x},t] = \int d^3\mathbf{x}\,n_{\text{pair}}$ is the total number of pairs produced which is independent of the observer position. So long as $\dot{N}$ maintains the same sign it follows that $F_\nu \leq F_0$. These formula confirm our rough numerology: a Crab nanoshot with $F_\nu = 100$ kJy ns would be caused by production of $N > 0.23\,\text{Cos}[\theta]$ mole of pairs. The sign of $\dot{N}$ can change if there is a field reversal, and both will happen in some cases [8]. If charge reversal occurs it will also result in additional coherent charge reversal EMP. Note that $F_\nu$ unaffected by dispersion along the line-of-sight can be considered an observable.

In terms of the distance $D$, the observable $F_\nu$ and $E_{\text{lim}}$ from Eqs. (2), (3) and (4) one can define estimators of the total pair production, total EMP energy, spark duration, EMP electric field, pair production rate, volume and a geometrical factor:

$$\hat{N} \equiv \sqrt{\frac{\pi c D^2}{e^2} F_0} = N\,\text{Sec}[\theta], \qquad \hat{\mathcal{E}} \equiv 4\pi D^2 \int_0^\infty d\nu\,F_\nu = 4\,\frac{e^2}{c}\int_{-\infty}^\infty dt\,\dot{N}^2\,\text{Sec}[\theta]^2,$$

$$\hat{\delta t} \equiv 4\,\frac{e^2}{c}\,\frac{\hat{N}^2}{\hat{\mathcal{E}}}, \qquad \hat{\delta E} \equiv \sqrt{\frac{8\pi\hat{\mathcal{E}}}{(c\,\hat{\delta t})^3}}, \qquad \hat{\Gamma} \equiv \frac{E_{\text{lim}}^2}{8\pi^2\,c\,e\,E_S\,\hat{\delta t}^2}, \qquad \hat{V} \equiv \frac{\hat{N}}{\hat{\delta t}\,\hat{\Gamma}}, \quad \text{and} \quad \hat{g} \equiv \frac{\hat{V}}{(c\,\hat{\delta t})^3}, \tag{9}$$

where $\hat{\delta t}$ and $\hat{\Gamma}$ can be identified as $\delta t_\Gamma$ and $\Gamma[E_{\text{lim}}]$, respectively. The relation of these estimators to physical properties of the spark does depend on the viewing angle, $\theta$, and the overall field geometry only some of which is encapsulated in $\hat{g}$, e.g. the observer frame is probably not a colinear frame and a boost corrections may be needed. For a fixed spark in a colinear frame $\hat{\mathcal{E}}, \hat{V} \propto \text{Sec}[\theta]^2$, $\hat{N}, \hat{\delta t} \propto \text{Sec}[\theta]$, $\hat{g}, \hat{\delta E} \propto \text{Cos}[\theta]$, and $\hat{\Gamma} \propto \text{Cos}[\theta]^2$. Nevertheless we do expect $\hat{g} \sim 1$ to be a rough distinguishing feature of Schwinger sparks. We do not but one could eliminate the explicit $\theta$ dependence of all these estimator by multiplying by appropriate powers of $\hat{g}$. All estimators can be expressed in terms of any two, e.g. we can estimate the peak EMP power in terms of $\hat{\delta t}$ and $\hat{g}$:

$$\frac{\hat{\mathcal{E}}}{\hat{\delta t}} = \frac{c(\hat{g}\,c\,\hat{\delta t}\,E_{\text{lim}}^2)^2}{(4\pi E_S)^2} = 1.30\left(\hat{g}\,\frac{\hat{\delta t}}{\text{nsec}}\right)^2 L_\odot.$$

**Schwinger EMP Characteristics**

When considering whether EMP from a surface spark could propagate out of an intervening magnetosphere one might first note that a Goldreich-Julian magnetosphere has particle density $n_{e^\pm} \geq \frac{2\pi}{ce}\Omega B$ so the plasma frequency is $\nu_{pl} = \sqrt{\frac{e^2 n_{e^\pm}}{\pi m_e}} \geq 250\,\text{GHz} \sqrt{\frac{B}{10^{13}\,\text{Gauss}}}$ which would not allow low amplitude GHz radiation to propagate. However nanoshot EMP has very large electromagnetic fields, with much larger energy density than that of the magnetospheric plasma. Furthermore nanoshots are temporally clustered in giant pulses (GPs) with duration $\sim 10\,\mu$sec (sometimes there are intermediate $\mu$sec timescale structures known as microbursts). As we now describe EMP is able to push plasma out of it's way at least temporarily. A $e^\pm$ which first encounters a pulse will be moved sideways by the transverse electric field, $E_{\text{EMP}}$, and then the $\boldsymbol{v} \times \boldsymbol{B}_{\text{EMP}}$ force will push it forward. Also acting is the neutron star's Lorenz forces $\boldsymbol{v} \times \boldsymbol{B}_{\text{NS}}$ which may push the particle sideways, out of the way of the EMP. This EMP particle acceleration will be a source of energy loss from the EMP and can degrade or effectively eliminate EMP from making it out of the pulsar magnetosphere. Even if individual nanoshots could not "blast" their way through a magnetosphere it is possible that trains of nanoshots could and even leave a temporary window for subsequent nanoshots to pass through. If a cluster of nanoshots could blast through for only a few $\mu$sec that would be sufficient to explain the observations even if temporally isolated nanoshots are not be able to propagate through the magnetosphere. This raises the question as to whether the observed temporal pulse structure has as much to do with the intervening plasma as with the sparks themselves. Curiously one naive magnetospheric recovery time, $10\,\text{km}/c \sim 30\,\mu$sec, is similar to the typical timescale of GPs. Perhaps these two timescales are related.

In this scenario it is clear that a large fraction of the EMP energy will go into exciting the magnetosphere, both in terms of energetic particles as well as magnetic field oscillations. The EMP that is observed has made it out of the magnetosphere and may be preferentially less effected by the intervening plasma. In any case we expect plasma effects to degrade emission at lower frequency more than at higher frequency. A net effect might be to decrease the value of the low frequency pulse parameter $F_0$ and increase $\hat{g}$ changing the scaling relation.

Note that Crab GPs and nanoshots are only observed in a narrow interval during the pulsar period. This would not be a property of a Schwinger spark since the EMP is not strongly beamed in any particular direction. We propose that the giant pulses and Schwinger sparks are occurring at other times but the geometry is not fortuitous for the EMP to make it to the Earth. The "observational window" for giant pulses could correspond to a "weak spot" in the magnetosphere perhaps when and where the line-of-sight from a hot spot to the Earth lines up with the magnetic field direction and/or where the Earth is nearly "overhead" of a hotspot providing a minimal column density to traverse. This can facilitate a blast through the magnetosphere. Furthermore our simple Schwinger spark model predicts linearly polarized pulses and not the diversity polarizations observed for nanoshots which are either linearly or circularly polarized in approximately equal proportions. Clearly further modeling of EMP propagation through a pulsar magnetosphere is required and may address these discrepancies.

Besides the charge separation EMP, secondary EMPs can be generated from Schwinger sparks. If $B$ is not uniform then the particle bunches will follow the field producing *coherent curvature* EMP. Also, *N*-suppressed sources of incoherent radiation, e.g. Larmor, synchrotron, spin-flip are allowed, in principle. The last two are further suppressed as the $e^\pm$ are created in the spin and synchrotron ground state (lowest Landau level) and can only emit synchrotron after propagating to a region where $B$ has changed amplitude or direction. The propagation of the EMP has been computed in vacuum and intervening plasma can change these results significantly.

The basic properties of EMPs from Schwinger sparks are general and possibly extended to other strong astrophysical emissions with different geometries. However one would not expect to find pulses with $\hat{g} \gg 1$ no matter the source of the EMP since this implies $\delta\hat{E} \gg 2.8 \times 10^{12}$ Gauss which the Schwinger mechanism would have acted to prevent, except in the improbable event that the pulse was a superposition of many short duration EMPs arriving simultaneously. More likely when a burst of radiation is a superposition of EMPs from many sparks they will not arrive simultaneously and $\hat{g}$ for the overall burst could be much less than unity. For example a Crab GP [3] might have 1 kJy mean flux at 10GHz over $10\,\mu$sec corresponding to $\hat{g} \sim 10^{-5}$. Fast Radio Bursts (FRBs) [10], if cosmological, emit $10^{40}$ erg over 5 msec also corresponding to $\hat{g} \sim 10^{-5}$. They could both be composed of a dilute superposition of EMPs from many Schwinger sparks.

### Ultra-Relativistic Electron Positron Beams

Because the primary source of ultra-relativistic $e^\pm$ is from vacuum pair production under a strong electric field,

each particle from the pairs produced will be accelerated to high energies along the electric field. The energy per particle may be estimated to be

$$\hat{\epsilon}_{\pm} \equiv e\, E_{\text{lim}}\, c\, \hat{\delta t} \approx 24.7\, \text{PeV}\, \frac{\hat{\delta t}}{\text{nsec}} \tag{10}$$

The total energy that goes into beams may be estimated as $\hat{\mathcal{E}}_{\pm} \equiv 2\hat{N}\hat{\epsilon}_{\pm} = \frac{4\pi}{\hat{g}}\frac{E_S}{E_{\text{lim}}}\hat{\mathcal{E}}$. Thus if Crab nanoshots are caused by Schwinger sparks then the observationally determined $\hat{\mathcal{E}}$ also gives us an estimate of the total energy dumped into PeV $e^{\pm}$.

A secondary source of ultra-relativistic $e^{\pm}$ will come from the EMP interaction with the pulsar magnetosphere which is believed to consist of an $e^{\pm}$ pair plasma. As already noted the outgoing EMP accelerates particles from the intervening plasma as it moves outward. Schwinger sparks produce strong EMP ($E_{\text{EMP}}\,\delta t \gg m_e\, c$) so the particles become relativistic and may become entrained in the pulse for a time which can be longer than $\delta t$, accelerating the particle to energies exceeding $E_{\text{EMP}}\, c\, \delta t$. As the propagating electric field falls off as $E_{\text{EMP}} \propto r^{-1}$ this entrainment will only increase the particle energies logarithmically and the maximum particle energies should be comparable to or less than those of particles accelerated in the spark. If the EMP is nearly completely degraded by interaction with the magnetosphere then the total energy in these secondary particle will be comparable to that in the primaries, although it will be distributed amongst a larger number of particles.

Since a Schwinger spark will produce very large numbers of $e^{\pm}$ pairs moving in opposite directions it can act like a linear $e^{\pm}$ accelerator. One might be interested in the annihilation products from collisions of a fraction of high-energy pairs within a spark producing a complete spectrum of standard model, and possibly beyond the standard model, particles. Also, the promising source of astrophysical observables may come from the $e^{\pm}$ beams after the spark. We have already mentioned possible coherent synchrotron emission. It is also likely that one of these beams would hit the neutron star itself which serves as a beam dump. While a neutron star can easily absorb a $\sim L_{\odot}/\text{m}^2$ beam without damage, this will temporarily heat a spot on the surface which will produce excess emission of photons and neutrinos. If the spark is near the surface the beam dump radiation may be temporally correlated with the EMP. Further modelling is required to predict the luminosity and spectrum (electromagnetic and neutrino) from these secondary processes.

## Summary

We have presented a model of Crab pulsar electromagnetic pulses called nanoshots which is clearly an extreme phenomena whose origin is currently not understood. Our model is based on breakdown of the vacuum in the presence of large electric fields, the Schwinger effects, whereby copious numbers of $e^{\pm}$ pairs are spontaneously produced, limiting the electric field to $E_{\text{lim}} \sim 2.5 \times 10^{12}$ Gauss in the colinear frame. The separation of the pairs in the remaining electric field produce a large EMP which is the proposed origin of the observed nanoshots. Produced pairs are accelerated to PeV energies and the energy loss to these $e^+$ and $e^-$ beams may lead to other observable follow on radiation. These sparks would most like be generated in small sub-meter scale regions near the neutron star surface and is clear that in this case the observed EMP is modified by passage through the pulsars magnetosphere. If this model can be corroborated then pulsar observations may be the first reasonably direct observation of the Schwinger effect and the nanoshots shed light on the sub-meter field structure of young pulsars which may be an important ingredient in the structure and evolution of young pulsars. Schwinger sparks may also play a role in bright emission from other bright radio transients such as fast radio bursts.

This is not the only explanation of GPs and nanoshots proposed so far (see [11-16]). Most of these models would have the nanoshots generated far from the surface, some as far as the light cylinder. There is phenomenology which may be difficult to understand with Schwinger sparks such as circular polarization [3] and banded emission in the spectra of interpulse (but not main pulse) GP's [11] but may be explained other model, e.g. [16]. The Schwinger spark numerology does provide a good fits to bright nanoshots which we find compeling although we would agree that the origin of GPs and nanoshots still remains an unsettled issue.

## Acknowledgments

This work was supported by the DOE at Fermi National Accelerator Laboratory under Contract No. DE-AC02-07CH11359. This work of HY was further supported by the DOE under Contract No. DE-FG02-95ER40896 and DE-AC02-05CH11231. The manuscript was partly written at the Aspen Center for Physics, supported in part by the NSF grant PHYS-1066293, while AS was attending the workshops "*Fast and Furious: Understanding Exotic Astrophysical Transcients*" and "*Ultra-compact Binaries as Laboratories for Fundamental Physics*". HY was supported by the Fermilab Fellowship in Theoretical Physics. The authors would like to thank Niyesh Afshordhi, Jon Arons, Latham Boyle, Ryan Chornock, Ue Li Pen, Peter Timbie and Neil Turok for useful discussions.**References**

A. [1] D. Staelin, E. Reifenstein *Pulsating Radio Sources near the Crab Nebula* Science **162** (1968) 1481
B. [2] T. Hankins, J. Kern, J. Weatherall, J. Eilek *Nanosecond radio bursts from strong plasma turbulence in the Crab pulsar* Nature. **422** (2003) 141
C. [3] A. Jessner *et al*. *Giant pulses with nanosecond time resolution detected from the Crab pulsar at 8.5 and 15.1 GHz* Asron. & Astrophys. **524** (2010) A60
D. [4] R. Manchester et al. *The Australia Telescope National Facility Pulars Catalog* Astron. J. **129** (1993) 4
E. [5] P. Goldreich, W. Julian *Pular Electrodynamics* Astrophys. J. **57** (1969) 869
F. [6] M. Baring, A. Harding *Photon splitting in soft gamma repeaters* Astrophys. J. **231** (1995) 77
G. [7] J. Schwinger *On Gauge Invariance and Vacuum Polarization* Phys. Rev. **82** (1951) 664
H. [8] R. Ruffini, G, Vereshchagin, S. Xue *Electron-Positron pairs in physics and astrophysics: From heavy nuclei to black holes* Phys. Rep. **487** (2010) 1
I. [9] A. Stebbins, H. Yoo, Classical Electrodynamics of Schwinger Sparks (2015) in preparation.
J. [10] D. Thornton *et al*. *A Population of Fast Radio Bursts at Cosmological Distances* Science **341** (2013) 53
K. [11] T. Hankins, J. Eilek *Radio Emission Signatures in the Crab Pulsar* Astrophys. J. **670** (2007) 693
L. [12] Mikhailovskii, O. Onishchenko, A. Smolyakov *An Interpretation of Pulsar Radio Micropulses* Sov. Astr. Lett. **11** (1985) 78
M. [13] J. Weatherall *Pulsar radio emission by conversion of plasma wave turbulence: Nanosecond time structure*. Astrophys. J. **506** (1998) 341
N. [14] S. Petrova *On the origin of giant pulses in radio pulsars* A&A **424** (2004) 227
O. [15] Y. Isotomin *Origin of Giant Pulses in Young Neutron Start and their Environments* IAU Symposium **218** (2004) 369
P. [16] M. Lyutikov *On the generation of Crab giant pulses* MNRAS **381** (2007) 1190
FERMILAB-PUB-15-236-A7